\documentclass[useAMS,usenatbib]{mn1e}
\usepackage{amssymb,amsmath,epsfig}
\allowdisplaybreaks

\title[Dynamical Instability of Cylinder]
{Dynamical Instability of Charged Gaseous Cylinder}
\author[M. Sharif and S. Mumtaz]{M. Sharif\thanks{E-mail:
msharif.math@pu.edu.pk (MS)} and Saadia Mumtaz\thanks
{sadiamumtaz17@gmail.com(SM)}\\
Department of Mathematics, University of the Punjab,\\
Quaid-e-Azam Campus, Lahore-54590, Pakistan.}
\begin{document}

\date{}

\volume{471}\pagerange{1215--1221} \pubyear{2017}

\maketitle

\label{firstpage}

\begin{abstract}
In this paper, we discuss dynamical instability of charged
dissipative cylinder under radial oscillations. For this purpose, we
follow the Eulerian and Lagrangian approaches to evaluate linearized
perturbed equation of motion. We formulate perturbed pressure in
terms of adiabatic index by applying the conservation of baryon
numbers. A variational principle is established to determine
characteristic frequencies of oscillation which define stability
criteria for gaseous cylinder. We compute the ranges of radii as
well as adiabatic index for both charged and uncharged cases in
Newtonian and post-Newtonian limits. We conclude that dynamical
instability occurs in the presence of charge if the gaseous cylinder
contracts to the radius $R_{*}$.
\end{abstract}


\section{Introduction}

A comprehensive study of collapsing systems and structure formation
of self-gravitating objects reveal interesting physical
perspectives. Charged self-gravitating objects may undergo various
evolutionary phases during gravitational collapse that results into
charged black holes or naked singularities. The stability of these
solutions under fluctuations has remarkable significance in general
relativity (GR). Initially, any stable system remains in state of
hydrostatic equilibrium unless its own gravity overcomes the
pressure which causes the matter to collapse. The collapsing system
contracts to a point under the influence of its own gravity leading
to compact objects.

The dynamical instability of massive stars can be studied in
Newtonian as well as post-Newtonian (pN) regimes \citep{a, a1}. This
provides a platform to evaluate ranges of deviation and level of
consistency between GR and Newton gravity. The analysis becomes
ambiguous in strong-field regimes due to non-linear terms, hence the
weak-field approximation schemes are used as an effective tool.
\citet{b} was the pioneer who discussed the concept of dynamical
instability of gaseous sphere by taking Newtonian perfect fluid in
terms of adiabatic index. He followed Eulerian approach for
linearized perturbed hydrodynamic equations and established a
variational principle to find characteristic frequencies in
Newtonian and pN limits. He also studied dynamical stability of
sphere under radial and non-radial oscillations at pN limit
\citep{c}.

\citet{d} investigated dynamical instability of spherical system
under perturbations by taking non-adiabatic fluid and found that the
instability range increases in Newtonian limit but decreases in pN
limit. Later, many researchers explored the influence of various
physical parameters on the dynamical instability of self-gravitating
systems under radial/non-radial perturbations \citep{e1, e2, e3}.
There has also been an extensive literature on the study of
cylindrical gravitational collapse with and without electromagnetic
field \citep{f1, f2, f3}. \citet{g} studied dynamical instability of
anisotropic collapsing cylinder in the context of expansion-free
model.

It is well-known that various physical aspects of matter
distribution play substantial role in the dynamical evolution of
self-gravitating systems. A star requires more electromagnetic
charge for its stability in a strong gravitational field. The
dynamical instability of collapsing systems in the presence of
electromagnetic field has a primordial history starting with
Rosseland \citep{1}. \citet{3} discussed the role of surface charge
in increasing stability of system with uniform density. \cite{4}
studied dynamical stability of sphere under radial pulsations in the
presence of electric charge. \citet{4a} found that neutron stars
having charge greater than the extreme value would explode.
\citet{5, 5a, 5b} studied the influence of electric charge on
dynamical instability of collapsing systems at Newtonian and pN
regimes.

In this paper, we study the impact of electromagnetic field on
dynamical instability of cylindrically symmetric collapsing system
by following Chandrasekhar approach \citep{b}. The format of the
paper is as follows. In section \textbf{2}, we provide some basic
equations and matter distribution for cylindrical geometry. Section
\textbf{3} deals with equations of motion under radial oscillations
following the Eulerian approach. We also formulate perturbed
pressure and adiabatic index in terms of Lagrangian displacement by
using conservation of baryon number. Section \textbf{4} is devoted
to find conditions for dynamical instability of homogeneous
cylinder. Finally, we conclude our results in the last section.

\section{Field Equations and Matter Configuration}

We consider a cylindrically symmetric system in the interior region
given by
\begin{equation}\label{5}
ds^2=-A^2(t,r)dt^2+B^2(t,r)dr^2+C^2(t,r)d\theta^2+dz^2,
\end{equation}
where the following restrictions on coordinates are taken to
preserve symmetry
\begin{equation}\nonumber
-\infty<t<\infty, \quad 0\leq r<\infty, \quad0\leq\theta\leq2\pi,
\quad-\infty<z<\infty.
\end{equation}
The corresponding Einstein field equations are given by
\begin{eqnarray}\label{6}
\frac{8\pi G}{c^4}T_{0}^0&=&\frac{1}{B^2}\left\{\frac{C''}{C
}-\frac{B'C'}{BC}\right\}-\frac{\dot{B}\dot{C}}{A^2BC},
\\\label{7}\frac{8\pi G}{c^4}T_{1}^1&=&\frac{1}{A^2}\left
\{\frac{\dot{A}\dot{C}}{AC}-\frac{\ddot{C}}{C}\right\}
+\frac{A'C'}{AB^2C},\\\label{7a}\frac{8\pi
G}{c^4}T_{2}^2&=&\frac{1}{AB}\left\{\frac{A''}{B}
-\frac{\ddot{B}}{A}+{\frac{\dot{A}\dot{B}}{A^2}
-\frac{A'B'}{B^2}}\right\},\\\nonumber\frac{8\pi
G}{c^4}T_{3}^3&=&\frac{A''}{AB^2}-\frac{\ddot{B}}{A^2B}
+\frac{\dot{A}\dot{B}}{A^3B}-\frac{A'B'}{AB^3}
+\frac{\dot{A}\dot{C}}{A^3C}-\frac{\ddot{C}}{A^2C}
\\\label{7b}&-&\frac{B'C'}{B^3C}+\frac{C''}{B^2C}+\frac{A'C'}{AB^2C}
-\frac{\dot{B}\dot{C}}{A^2BC},\\\label{8}\frac{8\pi
G}{c^4}T_{0}^1&=&\frac{1}{B^2}\left\{\frac{A'\dot{C}}{AC
}+\frac{\dot{B}C'}{BC}-\frac{\dot{C}'}{C}\right\},
\end{eqnarray}
where dot and prime denote derivatives with respect to $t$ and $r$,
respectively. The matter source is assumed to be locally charged
dissipative perfect fluid defined by
\begin{equation}\label{9}
T_{\nu}^\mu=(\sigma+p)u^\mu
u_{\nu}+p\delta_{\nu}^\mu+q^{\mu}u_{\nu}+q_{\nu}u^{\mu}
+\frac{1}{4\pi}[F_{\nu\rho}F^{\mu\rho}-
\frac{1}{4}\delta_{\nu}^{\mu}F_{\rho\lambda}F^{\rho\lambda}],
\end{equation}
where $p$ is the isotropic pressure, $\sigma$ is the energy density,
$F_{\mu\rho}$ is the Maxwell field tensor,
$u^{\mu}=\frac{dx^{\mu}}{ds}$ and $q^{\mu}$ represent four velocity
and radial heat flux, respectively satisfying $q_{\mu}u^{\mu}=0$.
Also, we have
\begin{equation}\nonumber
u^{\mu}=A^{-1}\delta^{\mu}_{0}, \quad q^{\mu}=q\delta^{\mu}_{1},
\quad u^{\mu}u_{\mu}=-1.
\end{equation}

We can define the electromagnetic field tensor in terms of four
potential as $F_{\mu\nu}=\Phi_{\nu;\mu}-\Phi_{\mu;\nu}$, which
satisfies the Maxwell field equations
\begin{equation}\nonumber
F^{\mu\nu}_{~;\nu}=4\pi J^\mu, \quad F_{[\mu\nu,\rho]}=0,
\end{equation}
where $J^\mu=\tilde{\rho} u^\mu$ is the four current. The
conservation equation, $J^\mu_{;\mu}=0$, yields
\begin{equation}\nonumber
Q(r)=4\pi\int_{0}^r\zeta BCdr,
\end{equation}
which is the total amount of charge within cylinder. We define the
electric field intensity as
\begin{equation}\label{9a}
E(t,r)=\frac{Q(r)}{2\pi C}.
\end{equation}
The conservation of energy-momentum tensor leads to the following
relations
\begin{eqnarray}\label{15}
\frac{\partial T_{0}^0}{\partial t}+\frac{\partial T_{1}^0}{\partial
r}+\frac{\dot{B}}{B}\left(T_{0}^0-
T_{1}^1\right)+T_{0}^1\left(\frac{B'}{B}+\frac{A'}{A}\right)=0,
\\\label{16}\frac{\partial T_{1}^0}{\partial
t}+\frac{\partial T_{1}^1}{\partial r}+\frac{A'}{A}\left(T_{1}^1-
T_{0}^0\right)+\left(\frac{\dot{A}}{A}+\frac{\dot{B}}{B}\right)T_{1}^0=0,
\end{eqnarray}
where $T_{0}^1=-\frac{A^2}{B^2}T_{1}^0$. The components of
energy-momentum tensor are
\begin{eqnarray}\nonumber
T^0_{0}=-\sigma+\frac{\pi}{2}E^2, \quad T^1_{1}=p+\frac{\pi}{2}E^2,
\quad T^2_{2}=T^3_{3}=p-\frac{\pi}{2}E^2.
\end{eqnarray}

In hydrostatic equilibrium, all the quantities governing motion
remain time independent. In this context, Eqs.(\ref{6}), (\ref{7})
and (\ref{16}) become
\begin{eqnarray}\label{19}
\frac{d}{dr}\left(\frac{C_{0}'}{B_{0}}\right)&=&\frac{8\pi G
}{c^4}B_{0}C_{0}\left(-\sigma_{0}+\frac{\pi E^2}{2}\right),
\\\label{20}\frac{dA_{0}}{dr}\frac{dC_{0}}{dr}&=&\frac{8\pi G
}{c^4}A_{0}B_{0}^2C_{0}\left(p_{0}+\frac{\pi E^2}{2}\right),
\\\label{21}(\sigma_{0}+p_{0})\frac{dA_{0}}{dr}&=&
-A_{0}\frac{d}{dr}\left(p_{0}+\frac{\pi E^2}{2}\right),
\end{eqnarray}
where zero suffix describes equilibrium state of the surface
stresses. We also have a useful relation through Eqs.(\ref{6}) and
(\ref{7}) given by
\begin{equation}\label{22}
\frac{8\pi
G}{c^4}(p_{0}+\sigma_{0})=\frac{1}{A_{0}C_{0}}\left\{\frac{1}{B_{0}^2}
\frac{dA_{0}}{dr}\frac{dC_{0}}{dr}\right\}-\frac{1}{B_{0}C_{0}}\left\{
\frac{d}{dr}\left(\frac{C_{0}'}{B_{0}}\right)\right\}.
\end{equation}
We take the exterior region for cylindrically symmetric spacetime in
retarded time coordinate $\nu$ defined as
\begin{eqnarray}\nonumber
ds^2&=&-\left(-\frac{2GM}{Rc^2}+\frac{GQ^2}{R^2c^4}\right)
d\nu^2-2d\nu dR\\\label{22a}&+&R^2(d\theta^2+\alpha^2dz^2),
\end{eqnarray}
where $\alpha$ is an arbitrary constant and $M$ is the total mass.
We choose the Schwarzschild coordinate as $C=r$ \citep{aa}.
\citet{bb} defined C-energy for cylindrically symmetric spacetime in
the form of mass function given by
\begin{eqnarray}\label{22b}
m(r)=\frac{1}{8}\left[1-\frac{1}{B_{0}^2}\right]+2\pi^2rE^2.
\end{eqnarray}
Differentiating this equation and using Eq.(\ref{7}), we have
\begin{eqnarray}\label{22c}
\frac{dm}{dr}=\frac{2\pi
rG}{c^4}\sigma_{0}-\frac{r\pi^2GE^2}{c^4}+\frac{d}{dr}(2\pi^2rE^2),
\end{eqnarray}
whose integration leads to
\begin{eqnarray}\label{22e}
m(r)=\frac{2\pi G}{c^4}\int_{0}^{r}r\sigma_{0}dr-\frac{G}{4c^4}
\int_{0}^{r}\frac{Q^2}{r}dr+\frac{Q^2}{2r}.
\end{eqnarray}
The equation for hydrostatic equilibrium can be obtained as
\begin{eqnarray}\label{22f}
\frac{dp_{0}}{dr}+\frac{G(8\pi
r^2p_{0}+Q^2)}{rc^4(1-8m+4Q^2)}-\frac{rQQ'-Q^2}{4\pi r^3}=0.
\end{eqnarray}

\section{Equations Governing Radial Oscillations}

In this section, we study dynamical characteristics of gaseous mass
undergoing radial oscillations. The non-vanishing components of four
velocity can be written as
\begin{equation}\label{25}
u^0=\frac{1}{A_{0}}, \quad u_{0}=-A_{0},\quad u^1=\frac{v}{A_{0}},
\quad u_{1}=\frac{B_{0}^2}{A_{0}}v,
\end{equation}
where $v=\frac{dr}{dt}$ corresponds to the radial velocity
component. We can evaluate these components with respect to
spacetime coordinates by taking $u^i=\frac{dx^i}{ds}$. We perturb an
equilibrium configuration in such a way that its cylindrical
symmetry does not change. The perturbed state with linear terms
yields
\begin{eqnarray}\nonumber
A&=&A_{0}+\delta A,\quad B=B_{0}+\delta B, \quad p=p_{0}+\delta p,
\quad \sigma=\sigma_{0}+\delta\sigma,
\\\label{25aa}Q&=&Q_{0}+\delta Q, \quad q=q_{0}+\delta q.
\end{eqnarray}
We apply the Eulerian approach \citep{c} for perturbations through
which the corresponding linearized forms (governing the radial
perturbations) of Eqs.(\ref{19}) and (\ref{20}) turn out to be
\begin{eqnarray}\label{30}
\frac{1}{r}\frac{\partial}{\partial r}\left(\frac{\delta
B}{B^3_{0}}\right)&=&\frac{8\pi
G}{c^4}\left(\delta\sigma-\frac{Q_{0}\delta Q}{4\pi r^2}\right),
\\\nonumber
\frac{8\pi G}{c^4}\left(\delta p+\frac{Q_{0}\delta Q}{4\pi
r^2}\right)&=&\frac{1}{rA_{0}B^2_{0}}\frac{\partial}{\partial
r}\left(\frac{\partial}{\partial r}\delta A-\frac{2\delta
B}{B_{0}}\frac{dA_{0}}{dr}\right),\\\label{31}
\end{eqnarray}
where $\delta A$, $\delta B$, $\delta\sigma$, $\delta p$ and $\delta
Q$ define the Eulerian changes. The linearized form of Eqs.(\ref{8})
and (\ref{16}) can be appropriately written as
\begin{eqnarray}\label{32}
&&\frac{1}{rB_{0}^3}\frac{\partial}{\partial t}\delta B=-\frac{8\pi
G}{c^4}\left[(p_{0}+\sigma_{0})v+\delta q\right],\\\nonumber
&&(p_{0}+\sigma_{0})\left(\frac{B_{0}}{A_{0}}\right)^2\frac{\partial
v}{\partial t}+\frac{\partial}{\partial r}\delta
p+\frac{1}{A_{0}}(p_{0}+\sigma_{0})\frac{\partial}{\partial r}\delta
A\\\nonumber&&+\frac{1}{A_{0}}(\delta
p+\delta\sigma)\frac{dA_{0}}{dr}+\frac{1}{4\pi
r^2}\frac{\partial}{\partial r}(Q_{0}\delta Q)-\frac{1}{2\pi
r^3}Q_{0}\delta Q \\\nonumber&&+\left[(p_{0}+\sigma_{0})
v-q_{0}\right]\left(\frac{B_{0}}{A_{0}}\right)^2\left
[\frac{1}{A_{0}}\frac{\partial}{\partial t}\delta
A+\frac{1}{B_{0}}\frac{\partial}{\partial t}\delta
B\right]=0.\\\label{33}
\end{eqnarray}

Let us introduce a Lagrangian displacement $``\eta"$ such that
$v=\frac{\partial\eta}{\partial t}$. Integration of Eq.(\ref{32})
gives
\begin{equation}\label{35}
\frac{1}{B_{0}^3r}\delta B=-\frac{8\pi
G}{c^4}(p_{0}+\sigma_{0})\eta+\int\delta qdt,
\end{equation}
which leads to
\begin{equation}\label{36}
-\frac{1}{B_{0}}\delta B=\frac{1}{A_{0}}\frac{dA_{0}}{dr}
+\frac{1}{B_{0}}\frac{dB_{0}}{dr}.
\end{equation}
Solving Eqs.(\ref{30}) and (\ref{35}), we have
\begin{equation}\label{38}
\delta\sigma=-\eta\frac{d\sigma_{0}}
{dr}-\eta\frac{dp_{0}}{dr}-\frac{1}{r}(p_{0}+\sigma_{0})\frac{\partial}
{\partial r}(r\eta)+\frac{Q_{0}}{4\pi r^2}\delta Q,
\end{equation}
which, in accordance with Eq.(\ref{21}), yields
\begin{eqnarray}\nonumber
\delta\sigma&=&-\eta\frac{d\sigma_{0}}{dr}-\frac{A_{0}}
{r}(p_{0}+\sigma_{0})\frac{\partial}{\partial
r}\left[\frac{r\eta}{A_{0}}\right]-\frac{\eta}{8\pi}\frac{d}{d
r}\left[\frac{Q^2}{r^4}\right]\\\label{39}&+&\frac{Q_{0}}{4\pi
r^2}\delta Q.
\end{eqnarray}
Substituting $\delta B$ from Eq.(\ref{35}) in (\ref{31}), we have
\begin{eqnarray}\nonumber
\frac{1}{rA_{0}B_{0}^2}\frac{\partial}{\partial r}\delta
A&=&\frac{8\pi G}{c^4}\left[\delta
p-\frac{2(p_{0}+\sigma_{0})\eta}{A_{0}}\frac{dA_{0}}{dr}
\right]+\frac{2 GQ_{0}}{r^2c^4}\delta Q\\\label{40}&-&\frac{16\pi
G}{c^4A_{0}}\frac{dA_{0}}{dr}\int\delta q dt,
\end{eqnarray}
which, through Eq.(\ref{22}), becomes
\begin{eqnarray}\nonumber
(p_{0}+\sigma_{0})\frac{\partial}{\partial r}\delta
A&=&\frac{dA_{0}}{dr}+\frac{A_{0}}{B_{0}}\frac{dB_{0}}{dr}\left[\delta
p-\frac{2}{A_{0}}\frac{dA_{0}}{dr}\left\{(p_{0}+\sigma_{0})\eta
\right.\right.\\\label{41}&+&\left.\left.\int\delta
qdt\right\}+\frac{Q_{0}}{4\pi r^2}\delta Q\right].
\end{eqnarray}

Now we consider time dependent perturbations $\eta e^{i\omega t}$,
where $\omega$ and $\eta$ represent characteristic frequency and
Lagrangian displacement, respectively, which associate fluid
elements in equilibrium with the perturbed configuration. These
equations are time dependent due to their natural modes of
oscillations. We can rewrite Eq.(\ref{33}) by taking $\delta
A,~\delta B,$ $\delta p,~\delta\sigma$, $\delta q$ and $\delta Q$ as
time dependent amplitudes of the respective quantities as
\begin{eqnarray}\nonumber
&&\omega^2\eta(p_{0}+\sigma_{0})\left(\frac{B_{0}}{A_{0}}\right)^2
=\frac{d}{dr}\delta p\delta+\delta
p\left[\frac{2}{A_{0}}\frac{dA_{0}}{dr}+\frac{1}{B_{0}}
\frac{dB_{0}}{dr}\right]\\\nonumber
&+&\frac{1}{A_{0}}\delta\sigma\frac{dA_{0}}{dr}
-\frac{2}{A_{0}}\left[(p_{0}+\sigma_{0})\eta+\int
qdt\right]\left[\frac{1}{A_{0}}\frac{dA_{0}}{dr}+\frac{1}{B_{0}}
\frac{dB_{0}}{dr}\right]\\\nonumber&+&\frac{Q_{0} \delta Q}{4\pi
r^2}\left[\frac{1}{A_{0}}\frac{dA_{0}}{dr}+\frac{1}{B_{0}}
\frac{dB_{0}}{dr}-\frac{2}{r}\right]+[(\sigma_{0}+p_{0})v-q_{0}]
\\\label{43}&\times&\left(\frac{B_{0}}{A_{0}}\right)^2
\left[\frac{1}{A_{0}}\frac{\partial}{\partial t}\delta
A+\frac{1}{B_{0}}\frac{\partial}{\partial t}\delta B\right].
\end{eqnarray}

\subsection*{The Conservation of Baryon Number}

The study of perturbed pressure in terms of Lagrangian displacement
requires an additional assumption through which one can discuss
physical aspects of gaseous mass undergoing adiabatic radial
oscillations. In this context, the required assumption can be
justified by the conservation of baryon numbers as $(Nu^j)_{;j}=0$,
or
\begin{equation}\label{45}
\frac{\partial}{\partial x^j}(Nu^j)+Nu^j\frac{\partial}{\partial
x^j}\ln\sqrt{-g}=0,
\end{equation}
where $N$ is the baryon number per unit volume. It plays a
substantial role in the evolution of various cosmic models.
According to this law, the total number of particles will remain
conserved during the fluid flow. The change in particle numbers
occurs due to the loss or gain of net fluxes. We consider a fluid
which satisfies this identity. Equation (\ref{45}) through
(\ref{25}) gives
\begin{eqnarray}\nonumber
&&\frac{\partial}{\partial
t}\left(\frac{N}{A_{0}}\right)+\frac{\partial}{\partial
r}\left(\frac{Nv}{A_{0}}\right)+\frac{Nv}{A_{0}}\left[
\frac{1}{A_{0}}\frac{\partial A}{\partial
t}+\frac{1}{B_{0}}\frac{\partial B}{\partial t}\right]\\\label{46}&&
+\frac{Nv}{A_{0}}\left[ \frac{1}{A_{0}}\frac{\partial A}{\partial
r}+\frac{1}{B_{0}}\frac{\partial B}{\partial
r}+\frac{1}{r}\right]=0.
\end{eqnarray}
We take a perturbation of the form
\begin{equation}\label{47}
N=N_{0}(r)+\delta N(r,t),
\end{equation}
such that Eq.(\ref{46}) with linear terms in $v$ yields
\begin{eqnarray}\label{48}
\frac{1}{r^2}\frac{d}{d
r}(\frac{N_{0}r^2v}{A_{0}/2})+\frac{1}{A_{0}}\frac{\partial}{\partial
t}\delta N+\frac{N_{0}}{A_{0}B_{0}}\frac{\partial}{\partial t}\delta
B+\frac{N_{0}v}{A_{0}B_{0}}\frac{dB_{0}}{dr}=0,
\end{eqnarray}
whose integration leads to
\begin{equation}\label{49}
\frac{1}{A_{0}}\delta
N+\frac{1}{r^2}\frac{d}{dr}\left(\frac{N_{0}r^2\eta}{A_{0}}\right)
+\frac{N_{0}}{A_{0}B_{0}}\left[\delta
B+\eta\frac{dB_{0}}{dr}\right]=0.
\end{equation}
Using Eq.(\ref{36}), it follows that
\begin{equation}\label{51}
\delta
N=N_{0}\left[\frac{\eta}{A_{0}}\frac{dA_{0}}{dr}-rB_{0}^2\int\delta
qdt\right]-\eta\frac{dN_{0}}{dr}-\frac{N_{0}A_{0}}{r^2}
\frac{\partial}{\partial r}\left(\frac{r^2\eta}{A_{0}}\right)=0.
\end{equation}
We assume an equation of state of the form
\begin{equation}
N=N(\sigma,p).
\end{equation}
Using Eqs.(\ref{39}) and (\ref{51}), we have
\begin{equation}\label{53}
\delta p=-\eta\frac{dp_{0}}{dr}-p_{0}\Gamma\frac{A_{0}}{r}
\frac{\partial}{\partial r}\left(\frac{r\eta}{A_{0}}\right)+\beta,
\end{equation}
where
\begin{eqnarray}\nonumber
\beta &=&\frac{1}{\partial N/\partial
p}\left[\frac{1}{4\pi}\frac{\partial
N}{\partial\sigma}\left\{\frac{\eta}{2}\frac{d}{dr}
\left(\frac{Q^2}{r^4}-\frac{Q_{0}\delta
Q}{r^2}\right)\right\}\right.\\\nonumber&+&\left.N_{0}\left\{\frac{\eta}{A_{0}}
\frac{dA_{0}}{dr}-rB_{0}^2\int\delta qdt\right\}\right],
\end{eqnarray}
and $\Gamma$ represents the adiabatic index defined by
\begin{equation}\label{54}
\Gamma=\frac{1}{p(\partial N/\partial
p)}\left\{N-(\sigma+p)\frac{\partial N}{\partial\sigma}\right\},
\end{equation}
which estimates the fluid stiffness and describes the pressure and
density fluctuations.

\section{Pulsation Equation and Variational Principle}

The linear pulsation is related to different modes of perturbations
applied to equilibrium cylindrical configuration and their
oscillation frequencies. Inserting $\delta\sigma$ and $\delta p$ in
Eq.(\ref{43}), we have
\begin{eqnarray}\nonumber
&&\omega^2B_{0}^2(p_{0}+\sigma_{0})\eta
=-\frac{d}{dr}\left(\eta\frac{dp_{0}}{dr}\right)
-\eta\frac{dp_{0}}{dr}\left[\frac{2}{A_{0}}
\frac{dA_{0}}{dr}\right.\\\nonumber&&+\left.\frac{1}{B_{0}}
\frac{dB_{0}}{dr}\right]-\frac{1}{A_{0}}\frac{dA_{0}}{dr}
\left[2(p_{0}+\sigma_{0})\eta\left\{\frac{1}{A_{0}}
\frac{dA_{0}}{dr}\right.\right.\\\nonumber&&+\left.\left.
\frac{1}{B_{0}}\frac{dB_{0}}{dr}\right\}+\frac{1}{r}
\frac{\partial}{\partial
r}\{r(p_{0}+\sigma_{0})\eta\}\right]\\\nonumber&&-\frac{d}{dr}
\left(p_{0}\Gamma\frac{A_{0}}{r}\frac{\partial}{\partial
r}\left(\frac{\eta
r}{A_{0}}\right)+\beta\right)-\left[\frac{2}{A_{0}}
\frac{dA_{0}}{dr}+\frac{1}{B_{0}}\frac{dB_{0}}{dr}\right]
\\\nonumber&&\times\left[p_{0}\Gamma\frac{A_{0}}{r}
\frac{\partial}{\partial r}\left(\frac{\eta
r}{A_{0}}\right)+\beta\right]-\frac{2}{A_{0}}\frac{dA_{0}}{dr}
\left[\frac{1}{A_{0}}\frac{dA_{0}}{dr}\right.\\\nonumber&&+\left.
\frac{1}{B_{0}}\frac{dB_{0}}{dr}\right]\int\delta
qdt+ \frac{Q_{0}\delta Q}{4\pi r^2}\left[\frac{1}{A_{0}}
\frac{dA_{0}}{dr}+\frac{1}{B_{0}}
\frac{dB_{0}}{dr}-\frac{2}{r}\right].\\\label{55}
\end{eqnarray}
Substituting $\frac{dp_{0}}{dr}$ from Eq.(\ref{21}), this leads to
\begin{eqnarray}\nonumber
&&\omega^2B_{0}^2=\frac{1}{A_{0}}\left[\frac{d^2A_{0}}{dr^2}
-\frac{1}{B_{0}}\frac{dB_{0}}{dr}+\frac{1}{r}\frac{dA_{0}}{dr}\right]
\\\nonumber&&-\frac{\eta}{8\pi}\frac{d^2}{dr^2}\left(\frac{Q^2}{r^2}\right)
\left[\frac{2}{A_{0}}\frac{dA_{0}}{dr}\frac{1}{B_{0}}\frac{dB_{0}}{dr}\right]
+\frac{\eta}{8\pi}\frac{d}{dr}\left(\frac{Q^2}{r^2}\right)
\\\nonumber&&
-\frac{d}{dr}\left[p_{0}\Gamma\frac{A_{0}}{r}\frac{\partial}{\partial}
\left(\frac{\eta
r}{A_{0}}+\beta\right)\right]-\left[p_{0}\Gamma\frac{A_{0}}{r}
\frac{\partial}{\partial}\left(\frac{\eta
r}{A_{0}}+\beta\right)\right]\\\nonumber&&\times\left[\frac{2}{A_{0}}
\frac{dA_{0}}{dr}+\frac{1}{B_{0}}\frac{dB_{0}}{dr}\right]
-\frac{2}{A_{0}}\frac{dA_{0}}{dr}\tilde{q}\left[\frac{2}{A_{0}}
\frac{dA_{0}}{dr}+\frac{1}{B_{0}}\frac{dB_{0}}{dr}\right]
\\\label{56}&&-\frac{Q_{0}\delta Q}{4\pi r^2}\left[\frac{1}{A_{0}}
\frac{dA_{0}}{dr}+\frac{1}{B_{0}}\frac{dB_{0}}{dr}-\frac{2}{r}\right],
\end{eqnarray}
where $\int\delta qdt=\widetilde{q}$. Using Eqs.(\ref{8}) and
(\ref{21}), we have
\begin{eqnarray}\nonumber
&&\omega^2B_{0}^2(p_{0}+\sigma_{0})\eta =\frac{8\pi
G}{c^4}p_{0}B_{0}^2(p_{0}+\sigma_{0})\\\nonumber&&+\frac{\eta}{r}
\left[\frac{dp_{0}}{dr}+\frac{1}{8\pi}\frac{d}{dr}
\left(\frac{Q^2}{r^2}\right)\right]-\frac{\eta}{8\pi}
\frac{d^2}{dr^2}\left(\frac{Q^2}{r^2}\right)\\\nonumber&&
+\frac{\eta}{8\pi}\frac{d}{dr}\left(\frac{Q^2}{r^2}\right)
\left[\frac{2}{A_{0}}\frac{dA_{0}}{dr}+\frac{1}{B_{0}}
\frac{dB_{0}}{dr}\right]\\\nonumber&&-\frac{d}{dr}
\left[p_{0}\Gamma\frac{A_{0}}{r}\frac{\partial}{\partial
r}\left({\eta
r}{A_{0}}\right)+\beta\right]-\frac{2}{A_{0}}\frac{dA_{0}}{dr}
\tilde{q}\\\nonumber&&\times\left[\frac{1}{A_{0}}
\frac{dA_{0}}{dr}+\frac{1}{B_{0}}\frac{dB_{0}}{dr}\right]
\\\label{59}&&+\frac{Q_{0}\delta Q}{4\pi r^2}\left[\frac{1}{A_{0}}
\frac{dA_{0}}{dr}+\frac{1}{B_{0}}\frac{dB_{0}}{dr}-\frac{2}{r}\right],
\end{eqnarray}
which is the required pulsation equation satisfying the boundary
conditions
\begin{eqnarray}\nonumber
&&\eta=0, \quad r=0, \quad\delta p=0, \quad r=R.
\end{eqnarray}
Taking the product of pulsation equation with $\eta r^2A_{0}B_{0}$
and integrating over values of $r$, it yields a characteristic value
problem for $\omega^2$ as
\begin{eqnarray}\nonumber
&&\omega^2\int_{0}^R r^2\eta^2AB^3(p+\sigma)dr=\frac{8\pi
G}{c^4}\int_{0}^R p(p+\sigma)r^2\eta^2AB^3dr\\\nonumber&&+\int_{0}^R
r\eta^2AB\left[\frac{dp}{dr}+\frac{1}{8\pi}\frac{d}{dr}
\left(\frac{Q^2}{r^2}\right)\right]dr-\int_{0}^R
r^2\eta^2\frac{dp}{dr}\\\nonumber&&\times
AB\left(p\Gamma\frac{A}{r}\frac{d}{dr}\left(\frac{\eta
r}{A}\right)+\beta\right)dr-\int_{0}^R
r^2\eta^2AB\left[p\Gamma\frac{A}{r}\right.\\\nonumber&&\times\left.
\frac{d}{dr}\left(\frac{\eta
r}{A}\right)+\beta\right]\left[\frac{1}{A}\frac{dA}{dr}+\frac{1}{B}
\frac{dB}{dr}\right]dr-2\int_{0}^R
r^2B\eta\frac{dA}{dr}\\\nonumber&&\times\tilde{q}\left[\frac{1}{A}
\frac{dA}{dr}+\frac{1}{B}\frac{dB}{dr}\right]dr+\frac{1}{8\pi}\int_{0}^R
Q_{0}AB\eta\delta
Q\left[\frac{1}{A}\frac{dA}{dr}\right.\\\label{61}&&+\left.\frac{1}{B}
\frac{dB}{dr}-\frac{2}{r}\right]dr.
\end{eqnarray}
We can define the orthogonality relation associated with this
equation as
\begin{equation}
\int_{0}^R AB^3r^2(p+\sigma)\eta^{(i)}\eta^{(j)}=0, \quad (i\neq j),
\end{equation}
where $\eta^{(i)}$ and $\eta^{(j)}$ provide proper solutions
corresponding to different characteristic values of $\omega^2$. The
study of dynamical instability of a star requires that the
right-hand side of Eq.(\ref{61}) must vanish by choosing a trial
function $\xi$ that satisfies the given boundary conditions.

In the following, we evaluate conditions for dynamical instability
by taking a homogeneous model.

\subsection*{The Homogeneous Model of Cylinder}

We study the conditions for dynamical instability of a homogeneous
cylinder with constant energy density. Equations (\ref{22e}) and
(\ref{22f}) governing the hydrostatic equilibrium allow the
integration such that we can write \citep{b}
\begin{equation}
y^2=1-\frac{r}{a^2}+\frac{b^2}{r^2}, \quad
y_{1}^2=1-\frac{R}{a^2}+\frac{b^2}{R^2},
\end{equation}
where $a^2=\frac{c^4}{2\pi G\sigma}$ and
$b^2=\frac{GQ^2(1-2c^2)}{2c^4}$. We can determine solutions of the
relevant physical quantities in terms of $y$ and $y_{1}$ as
\begin{eqnarray}\label{65a}
p=\sigma\frac{y-y_{1}}{3y_{1}-y}, \quad A^2=\frac{1}{4}[3y_{1}-y]^2,
\quad B^2=\frac{1}{y^2}.
\end{eqnarray}
For positivity of pressure, we have $3y_{1}>1$ which yields
\begin{equation}\nonumber
\frac{R}{a^2}-\frac{b^2}{R^2}<\frac{1}{9}.
\end{equation}
Using the inertial mass, this leads to
\begin{equation}\label{69}
R>9\left(\frac{2GM}{c^2}-\frac{GQ^2}{Rc^4}\right)=9R_{*},
\end{equation}
where $R_{*}$ is the limiting radius for charged cylinder. Inserting
the above physical quantities in Eq.(\ref{61}), it follows that
\begin{eqnarray}\nonumber
&&2a\omega^2y_{1}\int_{0}^{\xi_{1}}\frac{\xi^2\eta^2}
{y^3}d\xi=6y_{1}\int_{0}^{\xi_{1}}
\frac{y-y_{1}}{y^3(3y_{1}-y)^2}\xi^2\eta^2d\xi
\\\nonumber&&+\frac{3}{2a}\int_{0}^{\xi_{1}}\frac{3y_{1}-y}{y}
\xi\eta^2\frac{d}{d\xi}\left[\frac{y-y_{1}}{3y_{1}-y}
+\frac{G}{3ac^4}\frac{d}{d\xi}\left(\frac{Q^2}{\xi^2}\right)\right]d\xi
\\\nonumber&&-\frac{1}{2}\int_{0}^{\xi_{1}}\eta\xi^2\frac{3y_{1}-y}{y}
\frac{d}{d\xi}\left[\frac{y-3y_{1}}{a^3\xi}
\Gamma\frac{\partial}{\partial\xi}\left(\frac{\eta\xi}{3y_{1}-y}\right)
+\frac{3c^4\beta}{8\pi
G}\right]d\xi\\\nonumber&&-\frac{a^2}{2}\int_{0}^{\xi_{1}}\xi^2\eta\frac{3y_{1}-y}{y}
\left[\frac{y-y_{1}}{a\xi}\Gamma
\frac{\partial}{\partial\xi}\left(\frac{\eta\xi}{3y_{1}-y}\right)
+\frac{3c^4\beta}{8\pi
G}\right]\\\nonumber&&\times\left[\frac{2}{3y_{1}-y}\frac{d}{d\xi}
\left(3y_{1}-y\right)+y\frac{d}{d\xi}\left(\frac{1}{y}\right)\right]d\xi
-\frac{3ac^4}{8\pi G}\int_{0}^{\xi_{1}}\frac{\eta\xi^2}{y}
\tilde{q}\\\nonumber&&\times\frac{d}{d\xi}(3y_{1}-y)\left[\frac{1}{3y_{1}-y}\frac{d}{d\xi}
(3y_{1}-y)+y\frac{d}{d\xi}\frac{1}{y}\right]d\xi\\\nonumber&&
+\frac{3c^4}{(8\pi)^2 G}\int_{0}^{\xi_{1}}Q_{0}\xi\delta
Q\frac{3y_{1}-y}{2y}\left[\frac{2}{3y_{1}-y}\frac{d}{d\xi}(3y_{1}-y)
+y\frac{d}{d\xi}\frac{1}{y}\right]d\xi,\\\label{70}&&
\end{eqnarray}
where $\xi=\frac{r}{a}$, $\xi_{1}=\frac{R}{a}-\frac{b}{R}$ and
$\Gamma$ is taken to be constant.

We consider a trial function
\begin{equation}
\eta=\xi A=\frac{1}{2}\xi(y_{1}-y),
\end{equation}
such that Eq.(\ref{70}) becomes
\begin{eqnarray}\nonumber
&&\frac{a\omega^2y_{1}}{2}\int_{0}^{\xi_{1}}
\frac{\xi^4(3y_{1}-y)^2}{y^3}d\xi=\frac{3y_{1}}{2a}
\\\nonumber&&\times\int_{0}^{\xi_{1}}\frac{\xi^4(y-y_{1})(3y_{1}-y)}{4y^3}d\xi
+\frac{3}{2a}\int_{0}^{\xi_{1}}\frac{\xi^3(3y_{1}-y)^3}{2y}
\\\nonumber&&\times\frac{d}{d\xi}\left[\frac{y-y_{1}}{3y_{1}-y}
+\frac{G}{3ac^4}\frac{d}{d\xi}\left(\frac{Q^2}{\xi^2}\right)\right]d\xi
-\frac{1}{4}\int_{0}^{\xi_{1}}\frac{\xi^3(3y_{1}-y)^2}{y}
\\\nonumber&&\times\frac{d}{d\xi}\left[\frac{y-3y_{1}}{a^3\xi}
\Gamma\frac{\partial}{\partial\xi}
\left(\frac{\xi^2}{2}\right)+\frac{3c^4\beta}{8\pi
G}\right]d\xi-\frac{a^2}{4}\int_{0}^{\xi_{1}}\xi^3\\\nonumber&&\times\frac{(3y_{1}-y)^2}{y}
\left[\frac{y-y_{1}}{a\xi}\Gamma\frac{\partial}{\partial\xi}
\left(\frac{\xi^2}{2}\right)+\frac{3c^4\beta}{8\pi
G}\right]\\\nonumber&&\times\left[\frac{2}{3y_{1}-y}
\frac{d}{d\xi}\left(3y_{1}-y\right)+y\frac{d}{d\xi}
\left(\frac{1}{y}\right)\right]d\xi\\\nonumber&&-\frac{3ac^4}{16\pi
G}\int_{0}^{\xi_{1}}\frac{\xi^3(3y_{1}-y)}{y}\tilde{q}
\frac{d}{d\xi}(3y_{1}-y)\\\nonumber&&\times\left[\frac{1}{3y_{1}-y}
\frac{d}{d\xi}(3y_{1}-y)+y\frac{d}{d\xi}\frac{1}{y}\right]d\xi
\\\nonumber&&+\frac{3c^4}{(16\pi)^2 G}\int_{0}^{\xi_{1}}Q_{0}\xi\delta
Q\frac{(3y_{1}-y)^2}{y}\\\label{73}&&\times\left[\frac{2}{3y_{1}-y}
\frac{d}{d\xi}(3y_{1}-y)+y\frac{d}{d\xi}\frac{1}{y}\right]d\xi.
\end{eqnarray}
Inserting $y=\cos\theta$ and $\xi=\sin\theta$ in the above equation,
we have
\begin{eqnarray}\nonumber
&&\frac{(a\omega)^{2}\cos\theta_{1}}{2}\int_{0}^{\theta_{1}}
\frac{\sin^{4}\theta}{\cos^{2}\theta}(3\cos^{2}\theta_{1}
-\cos\theta)^2d\theta=\frac{3\cos\theta_{1}}{2}\\\nonumber&&\times\int_{0}^{\theta_{1}}
\frac{\sin^{4}\theta}{\cos^{2}\theta}[4\cos\theta\cos\theta_{1}
-3\cos^{2}\theta_{1}\cos^{2}\theta]d\theta\\\nonumber&&+\frac{3}{4}
\int_{0}^{\theta_{1}}(3\cos\theta_{1}-\cos\theta)^3
\frac{\sin^{3}\theta}{\cos\theta}\frac{d}{d\theta}\left[\frac{\cos\theta
-\cos\theta_{1}}{3\cos\theta_{1}-\cos\theta}\right.\\\nonumber&&\left.
-\frac{2GQ^2}{3ac}\frac{1}{\sin^3\theta}\right]d\theta-\frac{a}{4}
\int_{0}^{\theta_{1}}(3\cos\theta_{1}-\cos\theta)^2
\frac{\sin^{3}\theta}{\cos\theta}\\\nonumber&&\times\frac{d}{d\theta}\left[\frac{\cos\theta_{1}
-\cos\theta}{a^3\sin^4\theta}\Gamma+\frac{3c^4\beta}{8\pi
G}\right]d\theta\\\nonumber&&-\frac{a^3}{4}\int_{0}^{\theta_{1}}(3\cos\theta_{1}
-\cos\theta)^2\sin^{3}\theta\left[\frac{\cos\theta-\cos\theta_{1}}{a}\Gamma
\right.\\\nonumber&&+\left.\frac{3c^4\beta}{8\pi G}\right]
\left[\frac{2\sin\theta}{\cos\theta(3\cos\theta_{1}
-\cos\theta)}+\tan\theta\sec\theta\right]d\theta\\\nonumber&&-\frac{3ac^4}{16\pi
G}\int_{0}^{\theta_{1}}\frac{\sin^{3}\theta}{\cos\theta}(3\cos\theta_{1}
-\cos\theta)\tilde{q}\frac{d}{d\theta}(3\cos\theta_{1}
-\cos\theta)\\\nonumber&&\times\left[\frac{\sin\theta}{\cos\theta(3\cos\theta_{1}-\cos\theta)}
+\tan\theta\sec\theta\right]d\theta+\frac{3c^4\beta}{(16\pi)^2G}\\\nonumber&&
\times\int_{0}^{\theta_{1}}Q_{0}\delta
Q\sin\theta(3\cos\theta_{1}-\cos\theta)^{2}\\\label{75}&&\times\left[\frac{2\sin\theta}
{\cos\theta(3\cos\theta_{1}-\cos\theta)}
+\tan\theta\sec\theta\right]d\theta,
\end{eqnarray}
where $\theta_{1}=\sin^{-1}\left(\frac{R}{a}-\frac{b}{R}\right)$. By
taking $\omega^2=0$ and solving the integrals, we find exact
condition for marginal stability. We evaluate the values of
$\Gamma_{c}$ for $\theta$ such that $\Gamma\leq\Gamma_{c}$ for the
existence of dynamical instability. We also consider Newtonian limit
which implies that the resulting criteria for marginal stability is
$\Gamma>-\frac{9}{8}-\frac{81Q^2}{4}$. We compute $\Gamma$ and radii
of marginal stability for homogeneous gaseous cylinder corresponding
to $Q=0.4$ and $q=0.5$ which exhibit finite values of $\Gamma$ in
Newtonian and pN limits. We note that $\frac{R}{R_{*}}$ remains
positive for $\Gamma>0$ showing marginal stability of gaseous
cylindrical model in pN limit. The respective results are given in
Table \textbf{1}.
\begin{table}
\textbf{Table 1:} \textbf{Adiabatic Index and Radii for Homogeneous
Cylinder}
\vspace{0.5cm}\centering\\
\begin{small}
\begin{tabular}{|c|c|c|c|}
\hline\textbf{$\theta_{1}$}&\textbf{$R/R_{*}$}&
$\Gamma_{c}$ for $Q=0.4$\\
\hline{$0^\textmd{o}$}&{$10.364$}&{-4.365}\\
\hline{$10^\textmd{o}$}&{33.163}&{$2.894\times10^7$}\\
\hline{$20^\textmd{o}$}&{8.549}&{$3\times10^7$}\\
\hline{$30^\textmd{o}$}&{4.000}&{23647.19}\\
\hline{$40^\textmd{o}$}&{2.4203}&{131557}\\
\hline{$50^\textmd{o}$}&{1.704}&{87550}\\
\hline{$60^\textmd{o}$}&{1.333}&{118265.5}\\
\hline
\end{tabular}
\end{small}
\end{table}

The perturbation diverges exponentially for $\omega^2<0$ which
yields either expansion or contraction showing dynamical instability
of stellar model. In Newtonian limit, we explore the ranges of
instability for both charged (Figure \textbf{1}) as well as
uncharged cylinder (Figure \textbf{2}). Since the radius of
stability is a factor of $R_{*}$, so physically interesting results
can be obtained if $\frac{R}{R_{*}}\geq0$. For charged cylinder, we
find unstable radii corresponding to smaller values of charge. The
system becomes stable as charge increases. In case of uncharged
cylinder, dynamical instability occurs for $\Gamma<-1.125$. It is
obvious from the graph that for $\Gamma>-\frac{9}{8}$, the resulting
radius of stability is greater than $R_{*}$.

We obtain the following condition for dynamical instability of
relativistic gaseous masses including charge as
$\theta_{1}\rightarrow 0$
\begin{equation}\label{78}
\Gamma+\frac{3}{4}\left(\frac{3}{2}+27Q^2\right)<\frac{57}{42}\theta_{1}^2
=\frac{57}{42}\left [\frac{R}{a^2}-\frac{b^2}{R^2}\right].
\end{equation}
We can write
\begin{equation}\label{80}
R<\frac{57}{42\left[\Gamma+\frac{3}{4}\left(\frac{3}{2}+27Q^2\right)
\right]}\left[\frac{2GM}{c^2}-\frac{GQ^2}{Rc^4}\right],
\end{equation}
which leads to
\begin{figure}\center
\epsfig{file=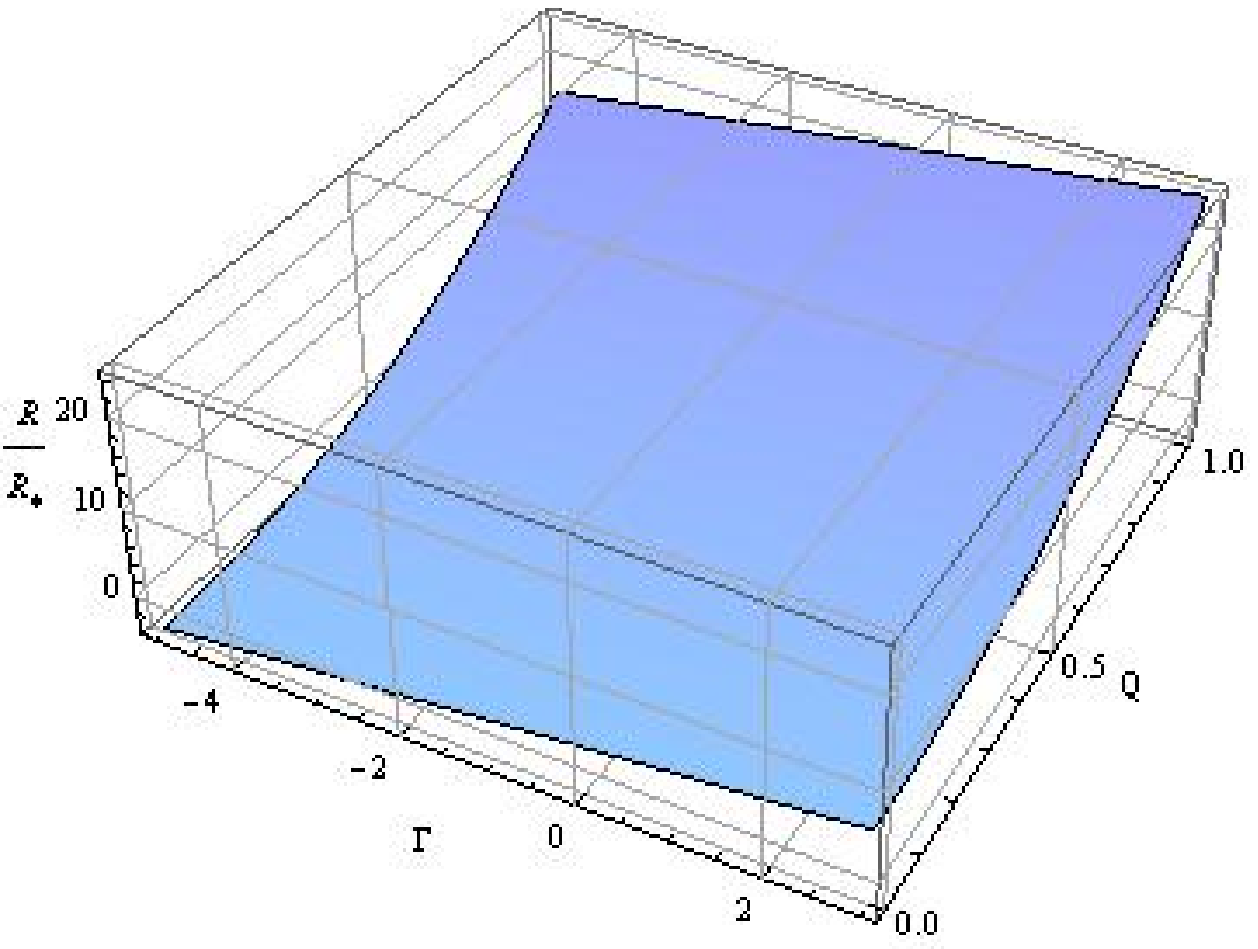,width=0.55\linewidth}\caption{Plot of
$\frac{R}{R_{*}}$ for dynamical stability/instability of charged
cylinder in Newtonian limit.}
\end{figure}
\begin{figure}\center
\epsfig{file=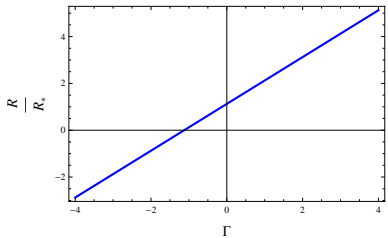,width=0.6\linewidth}\caption{Plot of
$\frac{R}{R_{*}}$ for dynamical stability/instability of uncharged
cylinder in Newtonian limit.}
\end{figure}
\begin{equation}\label{80}
\frac{R}{R_{*}}<\frac{K}{\left[\Gamma+\frac{3}{4}\left(\frac{3}{2}+27Q^2\right)
\right]},
\end{equation}
where $K=\frac{57}{42}$ for the homogeneous cylinder. This means
that if $\Gamma$ exceeds
$-\frac{3}{4}\left(\frac{3}{2}+27Q^2\right)$ by a small amount, the
dynamical instability can be prevented till the mass contracts to
radius $R_{*}$. The gaseous cylinder remains stable if its radius is
larger than $R_{*}$. The ranges of instability for charged
homogeneous cylindrical system are shown in Figure \textbf{3}. It
can be seen that the radius of stability is greater than $R_{*}$ for
$\Gamma>-1$ in pN limit. We also discuss the criteria and ranges of
instability for uncharged cylinder (Figure \textbf{4}). It is found
that $\frac{R}{R_{*}}\geq0$ when $\Gamma$ exceeds $-\frac{9}{8}$ by
a small amount showing stable cylindrical configuration. It is
observed that $\Gamma<-1.125$ leads to un-physical results as
$\frac{R}{R_{*}}<0$.
\begin{figure}\center
\epsfig{file=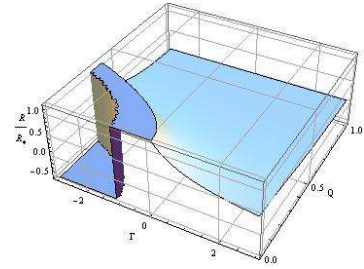,width=0.55\linewidth}\caption{Plot of
$\frac{R}{R_{*}}$ for dynamical stability/instability of charged
homogeneous cylinder.}
\end{figure}
\begin{figure}\center
\epsfig{file=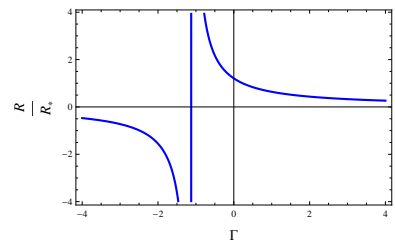,width=0.6\linewidth}\caption{Plot of
$\frac{R}{R_{*}}$ for dynamical stability/instability of homogeneous
uncharged cylinder.}
\end{figure}

\section{Outlook}

This paper is devoted to study the influence of electric charge on
dynamical instability of collapsing cylinder. We have followed
Eulerian and Lagrangian approaches to find linearized dynamical
equations as well as perturbed pressure. This perturbed pressure has
been obtained in terms of adiabatic index by taking conservation of
baryon numbers. A variational principle has been developed to
formulate characteristic frequencies of oscillation which refers to
the criteria of dynamical instability for gaseous cylinder. We have
also discussed conditions for dynamical instability by taking a
homogeneous model for cylinder.

We have computed particular values of radii as well as adiabatic
index $\Gamma$ to investigate the marginal stability of homogeneous
cylinder (Table \textbf{1}). It is found that $\Gamma$ takes finite
values greater than or equal to
$-\frac{3}{4}\left(\frac{3}{2}+27Q^2\right)$ for $Q=0.4$ and $q=0.5$
in Newtonian limit. In pN limit, $\frac{R}{R_{*}}$ remains positive
for $\Gamma>0$ showing marginal stability of gaseous cylindrical
model. We have also discussed the criteria for onset of dynamical
instability of gaseous masses.

In Newtonian limit, we have explored the ranges of instability for
both charged (Figure \textbf{1}) as well as uncharged cylinder
(Figure \textbf{2}). There is an extensive literature available for
dynamical instability of cylindrical gaseous systems using different
techniques in Newtonian limit. \citet{r1} studied dynamical
instability of self-gravitating cylindrical gaseous cloud by means
of normal mode analysis and found unstable solutions against various
types of perturbations. \citet{r2} discussed fragmentation of
cylindrical moleculer cloud with axial magnetic field on the basis
of a magnetohydrodynamical stability analysis and found that the
presence of magnetic field or rotation shortens the wavelength of
most unstable mode. \citet{r3} studied dynamical instability of a
self-gravitating magnetized cylindrical cloud by taking rotation
around its axis which suffers from various instabilities. \citet{r4}
explored dynamical instability of molecular cylindrical clouds
threaded by helical magnetic fields and found that all filamentary
molecular clouds initially in equilibrium state cannot be made to
undergo radial collapse by increasing the external pressure.
\citet{r5} discussed dynamical instability of cylindrical polytropic
filaments and found that the cylindrical polytropes converge at
large radii.

In our analysis, the gaseous cylinder remains stable as long as its
radius is larger than $R_{*}$ but becomes unstable as its radius
contracts to the radius $R_{*}$. For charged cylinder, dynamical
instability occurs for smaller values of charge whereas the system
becomes stable by increasing charge. The resulting radius of
stability is greater than $R_{*}$ for $\Gamma>-\frac{9}{8}$ in case
of uncharged cylinder while the dynamical instability occurs for
$\Gamma<-1.125$. It is mentioned here that electric charge plays a
substantial role to increase stability of cylindrical system as the
gaseous mass is more stable in Newtonian limit for larger values of
charge.

In pN limit, the gaseous cylinder undergoes dynamical instability if
$\Gamma$ exceeds $-\frac{3}{4}\left(\frac{3}{2}+27Q^2\right)$ by a
small amount. It is found that $\Gamma>-1$ and $Q>0.3$ provide valid
ranges of radii for the stability of cylinder whereas only unstable
radii exist corresponding to $\Gamma<-1$ and $Q<0.3$ (Figure
\textbf{3}). There is no effect of dissipation on stability of
collapsing system in this case. It is worth mentioning here that the
gaseous cylinder becomes unstable forever for smaller values of
charge. For uncharged cylinder, we have found that $\Gamma$ exceeds
$-\frac{9}{8}$ by a small amount showing stable cylindrical
configuration (Figure \textbf{4}). It is observed that
$\frac{R}{R_{*}}<0$ for $\Gamma<-1.125$ leading to un-physical
results. It is mentioned here that the cylindrical system is more
stable in Newtonian limit (Figure \textbf{1}) for larger values of
charge as compared to the post-Newtonian limit (Figure \textbf{3}).
We conclude that the presence of electromagnetic field plays a
remarkable role in the emergence of stability of gaseous cylinder.

\bsp

\label{lastpage}

\end{document}